\documentclass[singlespacing]{elsart}

\usepackage{graphicx}

\usepackage{amssymb}
\journal{Current Topics in
Catalysis}
\begin{document}

\begin{frontmatter}

\title{Relationship between preexponent and distribution over activation barrier energies for
enzymatic reactions}

\author{A.E. Sitnitsky},
\ead{sitnitsky@mail.knc.ru}

\address{Kazan Institute of Biochemistry and Biophysics, P.O.B. 30, Kazan
420111, Russia. e-mail: sitnitsky@mail.knc.ru, Tel. 7-843-2319037, Fax. 7-843-2927347.}

\newpage
\begin{abstract}
A relationship between the preexponent of the rate constant and the distribution over activation barrier energies for enzymatic/protein reactions is revealed. We consider an enzyme solution as an ensemble of individual molecules with different values of the activation barrier energy described by the distribution. From the solvent viscosity effect on the preexponent we derive the integral equation for the distribution  and find its approximate solution. Our approach enables us to attain a twofold purpose. On the one hand it yields a simple interpretation of the solvent viscosity dependence for enzymatic/protein reactions that requires neither a modification of the Kramers' theory nor that of the Stokes law. On the other hand our approach enables us to deduce the form of the distribution over activation barrier energies. The obtained function has a familiar bell-shaped form and is in qualitative agreement with the results of single enzyme kinetics measurements.
General formalism is exemplified by the analysis of literature experimental data.
\end{abstract}

\end{frontmatter}

\section{Introduction}
The idea of conformational heterogeneity (multiple conformational substates)
\cite{Aus75}, \cite{Ans85}, \cite{Fra91}, \cite{Fra01} was provided by new impetus
in the last fifteen years with the introduction of the concepts of static \cite{Xue95},
quasistatic \cite{Xue06}
and dynamic disorder \cite{Lu98}, \cite{Kou05}, \cite{Min05}, \cite{Ris08}. The results of these
studies suggest that there is a broad distribution over
the values of the catalytic reaction rate constant $k_2$
for an ensemble
of enzymes and the rate of a single enzyme strongly fluctuates in time
\cite{Xue95}, \cite{Lu98}, \cite{Xue06}, \cite{Kou05}, \cite{Min05}, \cite{Ris08}.
The reason is that in different conformational substates occupied by an enzyme the latter has different catalytic efficiency.
However it is widely believed that both static and dynamic disorder of reaction rates
are essentially indistinguishable in the ensemble averaged experiments \cite{Lu98}. This means that the
the distribution can not manifest
itself in the ensemble averaged kinetic measurements and is noticeable only in
single molecule kinetics \cite{Xue95},
\cite{Xue06}, \cite{Kou05}, \cite{Min05}, \cite{Ris08}. The main reason for the appearance of the distribution is believed to be the fluctuations of the value of the electric field strength in the active sites  of the enzymes from the ensemble \cite{Pra07}. The constant electric fields in the enzyme active sites created by preorganized dipoles of protein structure are considered as the key factor for the stabilization of the transition state and consequently for the values of the activation barrier energies \cite{War78}, \cite{War84}, \cite{Ols06}, \cite{Ols006}, \cite{War06}. In other words they determine the
enzyme catalytic efficiency \cite{War78}, \cite{War84}, \cite{Ols06}, \cite{Ols006}, \cite{War06}. Thus the variability of the electric field strength in the enzyme active sites has to lead to a broad distribution over the values of activation barrier energy, i.e., to that of the catalytic reaction rate constant.

In our opinion the distribution over the values of the activation barrier energy can manifest
itself in the ensemble averaged kinetic measurements.
Indeed dispersed kinetics resulted from heterogeneity were revealed in the ensemble studies by Frauenfelder and coworkers on rebinding of CO to hemeproteins upon photodissociation  \cite{Aus75}, \cite{Fra99}.
The fluctuations of rate constants (for which recently the term "dynamic disorder" was coined \cite{Lu98}, \cite{Kou05}, \cite{Min05}, \cite{Ris08}) were also inferred from the ensemble studies \cite{Aus75}.
One can conclude that if the distribution over the activation barrier energies
exists it must somehow affect ensemble kinetics. The distribution acquires
a status of a unique characteristic of the system and is of primary
interest. The aim of the present paper is to show that the existence of the distribution is closely related to the effective preexponent of the enzymatic reaction rate constant. The main premise of the paper is as follows: the existence of the distribution is sufficient for the appearance of the experimentally observable dependence of the enzymatic reaction rate constant on solvent viscosity. The latter manifests itself in the preexponent. Thus we persist to attain a twofold purpose. On the one hand we show that the existence of the distribution enables us to interpret experimental data for the effect of solvent viscosity on enzymatic reactions. On the other hand these data enable us to deduce the form of the distribution that can be in principle compared with that obtained from the single enzyme kinetic measurements. Unfortunately this comparison can not be carried out directly. The reason is that the studied by now enzymes exhibiting the solvent viscosity dependence have no reporter groups for the single kinetics measurements. The latter are feasible only for the enzymes possessing a unique fluorescent active group like, as an example, famous flavin adenine dinucleotide in the cholesterol oxidase \cite{Lu98}. However for such enzymes the data on solvent viscosity dependence of the reaction rate constant are not available.

The viscosity dependence of enzymatic and protein (ligand binding/rebinding) reactions
has been known for a long time \cite{Gav78},
  \cite{Gav79}, \cite{Bee80}, \cite{Gav80}, \cite{Dos83}, \cite{Gav86},
 \cite{Fra88},  \cite{Dem89}, \cite{Ng91},
\cite{Ng911}, \cite{Gav94}, \cite{Dos94}, \cite{Yed95}, \cite{Bar95}, \cite{Oh97}, \cite{Kle98},
\cite{Fra99}, \cite{Lav99}, \cite{Uri03}, \cite{Lav06}. For such reactions the functional dependence of the
reaction rate constant for the rate
limiting stage $k$ on solvent viscosity $\eta$ has the form
\[
k\propto \frac{1}{\left (\eta/\eta_0 \right )^{\beta}}
\]
where $\eta_0$ is the viscosity of pure solvent (for water $\eta_0 = 1\ cP$ at room temperature) and
$ 0 < \beta < 1 $ (usually $\beta \approx 0.4 \div 0.8$). This dependence is experimentally verified
in the range of variation of solvent viscosity by two orders of magnitude $\eta < 100\ cP$.
Similar dependence also takes place for folding of proteins
(see \cite{Pab04}, \cite{Fra06}, \cite{Kum08} and refs. therein) and at the formation of protein structure
\cite{Jas01}. However we will not touch upon these processes in the present paper.

The famous transition state theory as applied to the rate constants of enzymatic reaction
deals predominantly with their free energies (see extensive review of Truhlar and coauthors \cite{Gao06} and refs. therein) but fails to shed light on the solvent viscosity dependence of their preexponents.
The main tool for describing the solvent viscosity effect on the reaction rate constant is a high friction limit (also called strong damping or overdamped regime) of the Kramers' theory  \cite{Han90} or its modifications \cite{Gro80}, \cite{Zwa92} combined with the Stokes law for the friction coefficient.
The effect of friction on the processes in biomolecules is investigated in \cite{Dry10}.
There are many approaches to interpret the effect of solvent viscosity on the enzymatic rate constant \cite{Gav79}, \cite{Dos83}, \cite{Sch88}, \cite{Yed95}, \cite{Bar95}, \cite{Kle98}, \cite{Sit08}, \cite{Sit10}. A brief survey of these approaches was presented in our previous paper \cite{Sit10}. The model developed in \cite{Sit10}
required neither modification of the Kramers' model nor that of the Stokes law. The main premise of our approach was that a realistic enzyme solution was actually an ensemble of individual molecules with different characteristics (conditions for the movement of the system along the reaction coordinate).
It was shown that the experimentally observed dependence (\ref{eq1}) could be obtained if we took into account heterogeneity of conditions in the ensemble.
The effective reaction rate constant was obtained by averaging of individual Kramers' rate constants over the distribution.
The aim of \cite{Sit10} was to show that the idea of heterogeneity
enabled one to resolve the problem of solvent viscosity effect on the enzymatic reactions
in a conceptually much more simple way than modification of either the Kramers' theory or that of the Stokes law. Namely the latter approach was previously used for interpretation of the effect under consideration.

In this connection we stress that the idea about the importance of nonhomogeneity of protein solution for all aspects concerning the viscosity dependent effects was introduced earlier in the papers \cite{Bar95}, \cite{Lav99}, \cite{Lav06}.
In these papers a useful notion of local microscopic viscosity was shown to provide quantitative description of the experimental data on translational and rotational diffusion of proteins. On this approach the role of hydrodynamic interactions is highlighted. Making use of the notion of hydrodynamic radius the authors of  \cite{Lav99}, \cite{Lav06} succeeded in taking into account not only the molecular weight of cosolvent molecules but also their shape and size. In particular they  obtained quantitative description of translational and rotational
protein dynamics in the presence of polymeric cosolvent.
However taking into account the nonhomogeneity of the bulk solution is not sufficient for interpretation of the experimental data on solvent viscosity dependence for enzymatic and protein reactions. For the latter the protein structural fluctuations are of utmost importance \cite{Lav99}. In this case the  analysis similar to that performed for the rotational diffusion is not possible in this case because the size of that part of the protein responsible for the fluctuating motion is not known \cite{Lav99}.
For enzymatic and protein reactions only the experimental data for the dependence of fractional exponents on cosolvent molecular weight are available \cite{Yed95}.

In the present paper we retain the main idea of the model \cite{Sit10} but suggest its different technical realization. Unfortunately the way by which the distribution for the ensemble of enzymes can be introduced is not unique but this ambiguity seems to be inevitable. In \cite{Sit10} we made use of the distribution over the weight with which the contribution from solvent viscosity is taken into account in the viscosity for the movement of the system along the reaction coordinate. Thus in fact we carried out the averaging of the preexponent in the effective reaction rate constant over the {\it preexponents} of Kramers' rates for individual samples from the ensemble of enzymes. In the present paper we explore another option and show that the peculiarities of enzyme kinetics enable us to make use of the distribution over the activation barrier energies. As was mentioned above the existence of the latter is convincingly supported by the experimental data on single enzyme kinetics measurements. Thus in the present case we carry out averaging of the preexponent in the effective reaction rate constant over the {\it exponents} of Kramers' rates for individual samples from the ensemble of enzymes. The advantage of this approach is that the concept of the distribution over the activation barrier energies is beyond question in the community of the researches. Thus its application to the problem of solvent viscosity effect on enzymatic reactions seems to provide intuitively more simple and comprehensible physical picture.
Unfortunately this physical simplicity requires somewhat higher mathematical complexity of the present model compared with the previous one.

In the present model as well as in our previous one \cite{Sit10} the experimental dependence of the fractional exponent $\beta$ on the characteristics of cosolvent molecules is incorporated as input information. As was mentioned above for the enzymatic and protein reactions only the experimental data for the dependence of fractional exponents on cosolvent molecular weight are available \cite{Yed95}. This fact
imposes a natural limitation on both models. In the present model likewise in that of
\cite{Sit10} only cosolvent molecular weights but not the shape and size of cosolvent molecules are taken into account.

The paper is organized as follows. In Sec. 2 the discrete averaging of the rate constant for the ensemble of enzymes is discussed. In Sec.3 the approximation used for transition from discrete averaging to the continuous one is discussed in details. In Sec. 4 the continuous version of the averaging within the framework of the Kramers' theory is considered. In Sec. 5 the obtained integral equation for the distribution over the activation barrier energies is analyzed. In Sec.6 the general formalism is applied to the analysis of experimental data.
In Sec. 7 the results are discussed and the conclusions are summarized.

\section{Discrete averaging of the rate constant for enzymatic reactions}
Our primary aim is to derive the averaging procedure for the reaction rate constant.
First we consider the value of the reaction rate constant obtainable in the experiment.
The initial reaction rate for the  Michaelis-Menten
scheme (MM)
 $E+S$ $ \buildrel k_1 \over
\rightleftharpoons \atop k_{-1} $ $ES  \buildrel k_2 \over \rightarrow P+E$
 is given by the well known expression \cite{Dix79}
\begin{equation}
\label{eq1} V\equiv \frac{d\left[P\right]}{dt}=
\frac{k_2\left[E_T\right]\left[S\right]}
{K_M+\left[S\right]}
\end{equation}
where $\left[E_T\right]$ is the total substrate concentration and the Michaelis constant is
\begin{equation}
\label{eq2} K_M=\frac{k_{-1}+k_2}{k_1}
\end{equation}
The reaction rate constant is measured in excess of the substrate
\begin{equation}
\label{eq3}  \left[S\right]>> K_M
\end{equation}
Thus one obtains
\begin{equation}
\label{eq4} V\approx
k_2\left[E_T\right]
\end{equation}
Here $k_2$ is the effective reaction rate constant that is the average over the ensemble of enzymes. Thus in the experiment one actually measures the value
\begin{equation}
\label{eq5} k_2^{eff}\approx \frac{V}{\left[E_T\right]}
\end{equation}

Now let us derive the explicit expression for the value of $k_2^{eff}$. For the sake of convenience of the notation in this Sec. we assume that the ensemble of enzymes consists of a discrete set of pools with different values of the reaction rate constants $k_2^i$. We consider the simplest case of non-interacting channels. That is we consider the kinetic scheme  $E+S$ $ \buildrel k_1 \over
\rightleftharpoons \atop k_{-1} $ $ES^i  \buildrel k_2^i \over \rightarrow P+E$
where $i=1,2,3,...$. For the quasi-stationary concentration of the $i-$th enzyme-substrate complex ($d\left[ES^i\right]/dt=0$) we have the usual expression
\begin{equation}
\label{eq6} \left[ES^i\right]=\frac{k_1\left[E\right]\left[S\right]}
{k_{-1}+k_2^i}
\end{equation}
The total enzyme concentration is given by the expression
\begin{equation}
\label{eq7} \left[E_T\right]=\left[E\right]+\sum_{i}\left[ES^i\right]
\end{equation}
From (\ref{eq6}) and (\ref{eq7}) we obtain the quasi-stationary concentration of the $i-$th enzyme-substrate complex
\begin{equation}
\label{eq8} \left[ES^i\right]=\frac{k_1\left[S\right]\left[E_T\right]}
{\left(k_{-1}+k_2^i\right)\left\lbrace 1+
k_1\left[S\right]\sum_{j}1/\left(k_{-1}+k_2^j\right)\right\rbrace}
\end{equation}
Substituting (\ref{eq8}) into the expression for the initial reaction rate
\begin{equation}
\label{eq9} V\equiv \frac{d\left[P\right]}{dt}=
\sum_{i} k_2^i\left[ES^i\right]
\end{equation}
we obtain
\begin{equation}
\label{eq10} V=k_1\left[S\right]\left[E_T\right]
\sum_{i}\frac{k_2^i}{\left(k_{-1}+k_2^i\right)\left\lbrace 1+
k_1\left[S\right]\sum_{j}1/\left(k_{-1}+k_2^j\right)\right\rbrace}
\end{equation}
The requirement that the substrate is in excess takes the form
\begin{equation}
\label{eq11} k_1\left[S\right]\sum_{j}1/\left(k_{-1}+k_2^j\right) >> 1
\end{equation}
At the specified requirement we obtain
\begin{equation}
\label{eq12} V \approx \left[E_T\right]\sum_{i}k_2^i \delta_i
\end{equation}
where we have denoted
\begin{equation}
\label{eq13} \delta_i=\frac{1}{\left(k_{-1}+k_2^i\right)
\sum_{j}1/\left(k_{-1}+k_2^j\right)}
\end{equation}
It is obvious that the values $\delta_i$ represent normalized fractions, i.e.,
\begin{equation}
\label{eq14} \sum_{i}\delta_i=1
\end{equation}
Comparing (\ref{eq12}) with (\ref{eq5})  we finally obtain
\begin{equation}
\label{eq15} k_2^{eff}=\sum_{i}k_2^i \delta_i =
\left[\sum_{j}1/\left(k_{-1}+k_2^j\right)\right]^{-1}
\sum_{i}1/\left(1+k_{-1}/k_2^i\right)
\end{equation}
The latter is the discrete version of the averaging of the reaction rate constant for enzymatic reactions.

\section{Transition to continuous averaging for enzymatic reactions}
The dependence of $\delta_i$ on $k_2^i$ itself (see (\ref{eq13})) at discrete averaging of $k_2^i$ in (\ref{eq15}) is the main obstacle for direct generalization of the above described procedure to the continuous case. To overcome it we resort to the following approximation
\[
k_2^{eff}=\left[\sum_{j}1/\left(k_{-1}+k_2^j\right)\right]^{-1}
\sum_{i}1/\left(1+k_{-1}/k_2^i\right)\approx
\]
\begin{equation}
\label{eq16} \left[\sum_{j}\delta_j/\left(k_{-1}+k_2^j\right)\right]^{-1}
\sum_{i}\delta_i/\left(1+k_{-1}/k_2^i\right)
\end{equation}
The latter is the main approximation that we use in the present paper.
This approximation is motivated by the the fact that due to peculiarities of the enzyme kinetics the value of $\delta_i$ at averaging of $k_2^i$ in
(\ref{eq15}) is virtually a function of $k_2^i$ itself. The validity of approximation (\ref{eq16}) is supported by the fact that we replace $1$ by $\delta_i$ both in the numerator and denominator of the ratio of the sums. Indeed we can consider two limiting cases. In the first case the probabilities of all channels of decay from the set of $N$ channels are equal, i.e., $\delta_i=1/N$. Then it is obvious that the approximation (\ref{eq16}) is valid because it leads to strict equality. In the second case we consider the extremely narrow distribution (only one channel of decay). In the left hand side we model this case by setting $N=1$. In the right hand side
$\delta_i$ is reduced to Kronecker symbol $\delta_i=\delta_{il}$ where $\delta_{il}=1$ at $i=l$ and $\delta_{il}=0$ at $i\not= l$. Again the approximation (\ref{eq16}) is valid because it leads to strict equality. These arguments seem to be convincing enough to validate our approximation (\ref{eq16}) for enzymatic reactions.

\section{Continuous averaging for enzymatic reactions within the framework of Kramers' theory}
The main tool for describing the viscosity dependence of the reaction rate constant is a high friction limit
(also called strong damping or overdamped regime) of the Kramers' theory  \cite{Han90}. In it the reaction is conceived as a diffusion process of a particle with some effective mass along a reaction coordinate over some potential surface. The friction coefficient for the particle $\mu$ is supposed to obey the Stokes law $\mu =6\pi l \eta$ where $l$ is the characteristic linear size of the particle and $\eta$ is the viscosity for the movement of the system along the reaction coordinate. The discussion of the friction in biomolecules is presented in \cite{Dry10}. The famous Kramers' formula for the high friction limit is
\begin{equation}
\label{eq17} k_K=\frac{\omega_a\omega_b}{2\pi \mu}\exp\left[-\frac{\epsilon}{k_B T}\right]
\end{equation}
where $\omega_a$ and $\omega_b$ are characteristic frequencies of the potential surface at the bottom and at the top of the well respectively, $\epsilon$ is the activation barrier height, $k_B$ is the Boltzman constant and $T$ is the temperature.

We denote
\begin{equation}
\label{eq18} a=\frac{\omega_a\omega_b}{12\pi^2 l\eta_0}
\end{equation}
\begin{equation}
\label{eq19} \alpha=\frac{1}{k_B T}
\end{equation}
and rewrite the Kramers' formula as follows
\begin{equation}
\label{eq20} k_K=\frac{a}{\eta/\eta_0}\exp\left(-\alpha \epsilon\right)
\end{equation}
On the other hand as was discussed in Introduction for enzymatic reactions the experiment yields
\begin{equation}
\label{eq21} k_2^{eff}=\frac{b}{\left(\eta/\eta_0\right)^\beta}\exp\left(-\alpha E\right)
\end{equation}
Here $0 < \beta < 1$ is the empirical parameter while $b$ and $E$ are effective values for corresponding parameters. As was stated in Introduction the main premise of the present paper is to relate effective reaction rate constant $k_2^{eff}$ with the existence of some distribution $\rho (\epsilon)$ over the barrier heights $\epsilon$ in the ensemble of enzymes. In other words we assume that for an individual enzyme (sample) from the ensemble the Kramers' formula (\ref{eq20}) with a definite value of $\epsilon$ takes place. Then the effective rate constant (\ref{eq21}) results from the averaging over the ensemble with the distribution $\rho (\epsilon)$. Thus the ensemble is characterized by the distribution $\rho (\epsilon)$ over the values of the energy $\epsilon$. To stress the used approximation (\ref{eq16}) we further write the continuous distribution function explicitly as $\rho(\epsilon, k_2(\epsilon))$ when it is necessary.

At the specified approximation (\ref{eq16}) the continuous version of the discrete averaging considered in Sec.2 is obtained as follows:
\[
 k_2^i \longrightarrow k_2(\epsilon)=\frac{a}{\eta/\eta_0}\exp\left(-\alpha \epsilon\right)
\]
\[
\sum_{j}\delta_j/\left(k_{-1}+k_2^j\right)\longrightarrow
\int\limits_{0}^{\infty}d \epsilon\ \frac{\rho(\epsilon)}{k_{-1}+
a/\left(\eta/\eta_0\right)\exp\left(-\alpha \epsilon\right)}
\]
\[
\sum_{i}\delta_i/\left(1+k_{-1}/k_2^i\right)\longrightarrow
\int\limits_{0}^{\infty}d \epsilon\ \frac{\rho(\epsilon)}{1+
\left(k_{-1}/a\right)\left(\eta/\eta_0\right)\exp\left(\alpha \epsilon\right)}
\]
\begin{equation}
\label{eq22} \sum_{i}\delta_i=1\longrightarrow
\int\limits_{0}^{\infty}d \epsilon\ \rho(\epsilon)=1
\end{equation}
\begin{equation}
\label{eq23} E=\sum_{i}\epsilon_i\delta_i\longrightarrow
E=\int\limits_{0}^{\infty}d \epsilon\ \rho(\epsilon)\epsilon
\end{equation}
The strict relationship
\begin{equation}
\label{eq24} k_2^{eff}=\sum_{i}k_2^i \delta_i\longrightarrow
k_2^{eff}=\int\limits_{0}^{\infty}d \epsilon\ \rho\left(\epsilon, k_2(\epsilon)\right)k_2(\epsilon)
\end{equation}
is replaced with the help of the assumption (\ref{eq16}) by the approximate one
\[
k_2^{eff}=\sum_{i}k_2^i \delta_i\longrightarrow k_2^{eff}\approx
\]
\begin{equation}
\label{eq25}\int\limits_{0}^{\infty}d \epsilon\ \frac{\rho(\epsilon)}{1+
\left(k_{-1}/a\right)\left(\eta/\eta_0\right)\exp\left(\alpha \epsilon\right)}\left[\int\limits_{0}^{\infty}d \epsilon\ \frac{\rho(\epsilon)}{k_{-1}+
a/\left(\eta/\eta_0\right)\exp\left(-\alpha \epsilon\right)}\right]^{-1}
\end{equation}
Taking into account (\ref{eq21}) we obtain from
the latter expression the equation for the unknown function of the distribution
$\rho (\epsilon)$
\[
\frac{b}{\left(\eta/\eta_0\right)^\beta}\exp\left(-\alpha E\right)=
\]
\begin{equation}
\label{eq26}
\int\limits_{0}^{\infty}d \epsilon\ \frac{\rho(\epsilon)}{1+
\left(k_{-1}/a\right)\left(\eta/\eta_0\right)\exp\left(\alpha \epsilon\right)}\left[\int\limits_{0}^{\infty}d \epsilon\ \frac{\rho(\epsilon)}{k_{-1}+
a/\left(\eta/\eta_0\right)\exp\left(-\alpha \epsilon\right)}\right]^{-1}
\end{equation}
We denote
\begin{equation}
\label{eq27} c=\frac{k_{-1}}{a}\frac{\eta}{\eta_0}
\end{equation}
Then (\ref{eq26}) can be cast in the form
\begin{equation}
\label{eq28} \exp\left(-\alpha E\right)\frac{b}{a}\left(\frac{\eta}{\eta_0}\right)^{1-\beta}
\int\limits_{0}^{\infty}d \epsilon\ \frac{\rho(\epsilon)\exp\left(\alpha \epsilon\right)}
{1+c\exp\left(\alpha \epsilon\right)}=
\int\limits_{0}^{\infty}d \epsilon\ \frac{\rho(\epsilon)}
{1+c\exp\left(\alpha \epsilon\right)}
\end{equation}
Taking into account the obvious identity
\begin{equation}
\label{eq29} \frac{\exp\left(\alpha \epsilon\right)}
{1+c\exp\left(\alpha \epsilon\right)}=\frac{1}{c}\left(1-\frac{1}{1+c\exp\left(\alpha \epsilon\right)}\right)
\end{equation}
and the righthand side of (\ref{eq22}) we obtain the integral equation for the distribution $\rho (\epsilon)$
\begin{equation}
\label{eq30} \int\limits_{0}^{\infty}d \epsilon\ \frac{\rho(\epsilon)}
{1+\left(k_{-1}/a\right)\left(\eta/\eta_0\right)\exp\left(\alpha \epsilon\right)}=
\frac{1}{1+\left(k_{-1}/b\right)\left(\eta/\eta_0\right)^\beta\exp\left(\alpha E\right)}
\end{equation}
The righthand sides of (\ref{eq22}) and (\ref{eq23}) yield two equations
\begin{equation}
\label{eq31}
\int\limits_{0}^{\infty}d \epsilon\ \rho(\epsilon)=1
\end{equation}
\begin{equation}
\label{eq32}
E=\int\limits_{0}^{\infty}d \epsilon\ \rho(\epsilon)\epsilon
\end{equation}
for finding two effective parameters $b$ and $E$ that can be directly compared with the experimentally observable values (note that the parameter $a$ is expressed via molecular parameters by (\ref{eq18}) and is considered as a given one). Thus the equations (\ref{eq30}), (\ref{eq31}) and (\ref{eq32}) represent a closed system of equations for our problem. It is worthy to stress that the equation (\ref{eq30}) (provided that the approximation (\ref{eq16})) is applicable as long as the requirement that the substrate is in excess
\begin{equation}
\label{eq33}
 k_1\left[S\right]\int\limits_{0}^{\infty}d \epsilon\ \frac{\rho(\epsilon)}{k_{-1}+
a/\left(\eta/\eta_0\right)\exp\left(-\alpha \epsilon\right)} >> 1
\end{equation}
is satisfied.

\section{Analysis of the integral equation (\ref{eq30})}
We denote
\begin{equation}
\label{eq34}  q=\frac{k_{-1}\exp\left(\alpha E\right)}{b}\left(\frac{a}{k_{-1}}\right)^\beta
\end{equation}
As we will see later in practice the requirement
\begin{equation}
\label{eq35} q >> 1
\end{equation}
is satisfied. In this range the following solution of the integral equation (\ref{eq30}) can be guessed
\[
\rho(\epsilon)=\frac{\alpha \sin (\pi \beta)}{2\pi}\Biggl [
\frac{1}{\cos (\pi \beta)+\cosh[\alpha \beta \epsilon-\ln q]}+
\]
\begin{equation}
\label{eq36} \frac{\pi}{q}
\sum_{n=0}^{\left[1/\beta-1\right]}\left(-\frac{1}{q}\right)^n \exp\{-\left[1-(1+n)\beta\right]\alpha \epsilon \}\Biggr ]
\end{equation}
where $[x]$ in $\sum_{...}^{\left[x\right]}$ means the integer part of $x$.
The latter distribution can be verified by direct substitution into (\ref{eq30}) and numerical integration. It yields an excellent coincidence of the lefthand side of the equation with the function in its righthand side for a wide range of physically reasonable parameters outlined by the requirement (\ref{eq35}).
In Fig. 1 and  Fig. 2 the distribution (\ref{eq36}) is depicted at different combinations between the values of the parameters $\beta$ and $q$. In the limit of very large $q$ ($q >>> 1$) the distribution (\ref{eq36}) is transformed into the well known Cole-Cole one
\[
\rho(\epsilon)=\frac{\alpha \sin (\pi \beta)}{2\pi}\Biggl [
\frac{1}{\cos (\pi \beta)+\cosh[\alpha \beta \epsilon-\ln q]}\Biggr ]
\]

Within the range of validity of (\ref{eq36}) outlined by the requirement (\ref{eq35}) we obtain from (\ref{eq31})
\begin{equation}
\label{eq37} ctg \left[\pi \beta\left(1-\frac{\sin (\pi \beta)}{2q(1-\beta)}\right)+O\left(\frac{1}{q^2}\right)\right]=\frac{1+q\cos (\pi \beta)}{q\sin (\pi \beta)}
\end{equation}
From (\ref{eq32}) we obtain
\[
 \pi \alpha E \beta^2=\ln (q)\ \arctan \left(\frac{\sin (\pi \beta)}{1+\cos (\pi \beta)}\right)+2\pi \beta \ln 2-4L\left(\frac{\pi \beta}{2}\right)+
\]
\[
L\left(\theta-\frac{\pi (1- \beta)}{2}\right)-L\left(\theta+\frac{\pi (1- \beta)}{2}\right)+2L\left(\frac{\pi (1- \beta)}{2}\right)+\frac{\ln (q)}{2}\times
\]
\begin{equation}
\label{eq38} \left\lbrace\frac{\pi}{2}-
\arcsin\left[ \frac{1+\cosh \left(\ln (q)\right)\cos (\pi \beta)}{\cos (\pi \beta)+\cosh \left(\ln (q)\right)}\right]\right\rbrace + \frac{\pi \beta \sin (\pi \beta)}{2q\left(1-\beta\right)}+O\left(\frac{1}{q^2}\right)
\end{equation}
where
\begin{equation}
\label{eq39} \theta =\arctan\left[\tanh\left(\frac{\ln (q)}{2}\right)\tan\left(\frac{\pi \beta}{2}\right)\right]
\end{equation}
and $L(x)$ is the Lobachevskii function $L(x)=-\int\limits_{0}^{x}d t\ \ln \left(\cos t\right)$.
Recalling that $q=q(b,E)$ (see (\ref{eq34})) we conclude that equations (\ref{eq37}) and (\ref{eq38})
are a system of two equations for two unknown parameters $b$ and $E$.

There are two ways to analyze the equations obtained.
The first way can be called a direct (or purely theoretical) one. On this approach we assume the parameter $a$ given by (\ref{eq18}) as a known one (i.e., we assume $\omega_a$, $\omega_b$ and $l$ to be known values), calculate theoretical parameters $b$ and $E$ from (\ref{eq37}) and (\ref{eq38}), identify these parameters with the effective values of corresponding parameters known from the experiment and finally compare the theoretical values obtained with the experimental values. Parameter $q=q(b,E)$ is eliminated as an independent value on this approach.
In practice this way is hampered by the fact that the values $\omega_a$, $\omega_b$ and $l$ are generally unknown and besides it is rather difficult to solve the system of equations (\ref{eq37}) and (\ref{eq38}). The second way can be called an inverse (or experimentally motivated) one. On this approach we identify parameters $b$ and $E$ with the effective values extracted from the experiment,
consider them as given values while consider the parameters $a$ and $q$ as unknown values. The equation (\ref{eq37}) at large values of $q$ transforms into approximate identity. Thus we conclude that within the range of the validity of our approach outlined by the requirement (\ref{eq35}) the normalization of the distribution is always approximately satisfied. Then
equation (\ref{eq38}) gives the relationship between the parameters $q$ and $E$ (direct dependence $E=E(q)$ or implicit dependence $q=q(E)$). In Fig.3 this relationship is depicted at different values of the parameter $\beta$. At given values of $E$ and $\beta$ we can define from (\ref{eq38}) the value of the parameter $q$.
From (\ref{eq34}) we obtain that the parameter $a$ is unequivocally determined by $q$ at given values of $\beta$, $b$ and $E$
\begin{equation}
\label{eq40} a=k_{-1}\left[\frac{bq}{k_{-1}\exp\left(\alpha E\right)}\right]^{1/\beta}
\end{equation}

In practice it is more convenient to use the second way.
Thus the following strategy is suggested by the present approach. We consider the empirical fractional exponent $\beta$, the effective preexponent parameter $b$, the effective activation energy $E$ and the reaction rate constant $k_{-1}$ as the given experimental values. Then we solve the implicit equation (\ref{eq38}) to find the value of the parameter $q$ as a function of the values $\beta$ and $E$. If the obtained value of $q$ satisfies the requirement (\ref{eq35}) then our distribution (\ref{eq36}) satisfies the normalization requirement (\ref{eq31}).
Knowing $\beta$, $b$, $E$, $q$ and $k_{-1}$ we obtain from (\ref{eq40}) the value of the unknown parameter $a$. The latter is directly related to molecular parameters of the system via relationship (\ref{eq18}) and thus provides the information of major interest.
From consideration of (\ref{eq20}) and (\ref{eq21}) we intuitively anticipate that
the value of the parameter $a$ should not differ significantly from the value of the parameter $b$. Then the proximity of the obtained value of $a$ to the value of $b$ means that our model works satisfactorily well and our approach yields the reasonable interpretation of the experimental data.
In the next Sec. we exemplify this strategy by the analysis of data for oxygen escape from hemerythin presented in \cite{Yed95}.

\section{Analysis of the literature experimental data}
There are very few enzymatic/protein reactions for which a sufficiently complete set of the experimental data can be found in literature. In fact oxygen escape from hemerythin explored in \cite{Yed95} seems to be a unique example. Even for this protein "glycerol is the only cosolvent for which extensive data are available" \cite{Yed95}. For oxygen escape from hemerythin with viscosity varied by glycerol the rate constant is $k_{esc}\left(s^{-1}\right)=4\times 10^9 \left(\eta/\eta_0\right)^{-0.54}\exp \left[-H_{esc}/\left(RT\right)\right]$ where
$H_{esc}=13\ kJ \cdot mol^{-1}$ at $278\ K$ \cite{Yed95}.
Thus we have the following values of the parameters: $b=4\times 10^9$, $\beta=0.54$ and $E=13\ kJ \cdot mol^{-1}$ at $278\ K$ \cite{Yed95}. The latter means that $\alpha E=E/\left(k_B T\right)\approx 5.63$. From Fig. 3 we obtain for such values of $\alpha E$ and $\beta$ that $\log\ q \approx 1.2$, i.e., $q \approx 15.85$. This value of the parameter $q$ satisfies the requirement (\ref{eq35}) while with some strain (see Discussion), i.e., our distribution (\ref{eq36}) satisfies the normalization requirement (\ref{eq31}).
The distribution (\ref{eq36}) for these values of the parameters is depicted in Fig. 4. Both hands of the integral equation (\ref{eq30}) are plotted in Fig.5 for the distribution (\ref{eq36}) and the parameters obtained.

For oxygen escape from hemerythin the role of $k_{-1}$ is played by the rate constant
$k_{int}\left(s^{-1}\right)=10^8 \exp \left[-H_{int}/\left(RT\right)\right]$ where $H_{int}=4\ kJ \cdot mol^{-1}$ at $278\ K$ \cite{Yed95}. Thus we have $k_{-1}\approx 1.77 \cdot10^7\ s^{-1}$. Substituting all parameters into (\ref{eq40}) we finally obtain $a\approx 2.0\cdot 10^9$. This value closely matches the value of the parameter $b=4\times 10^9$ that provides strong evidence in favor of our approach. We conclude
that for oxygen escape from hemerythin the combination of molecular parameters is
\[
a=\frac{\omega_a\omega_b}{12\pi^2 l\eta_0}\approx 2.0\cdot 10^9
\]
The latter estimate enables one to make assumptions about the shape of the potential energy along the reaction coordinate (characterized by the frequencies $\omega_a$ and $\omega_b$) or the linear size $l$ of the the effective particle used in the Kramers' theory.

\section{Discussion}
We have revealed the relationship between the preexponent of the effective reaction rate constant and the distribution over the activation barrier energies for enzymatic/protein reactions. This relationship arises from the averaging of the reaction rate constant in enzyme solution over the ensemble of  individual enzyme molecules described in Sec. 4. It results from the fact that
due to peculiarities of the enzyme kinetics the distribution function $\rho(\epsilon, k_2(\epsilon))$ at averaging of $k_2(\epsilon)$ is virtually a function of $k_2(\epsilon)$ itself. Thus at averaging
\[
k_2^{eff}=\int\limits_{0}^{\infty}d \epsilon\ \rho\left(\epsilon, k_2(\epsilon)\right)k_2(\epsilon)
\]
the dependence $k_2(\epsilon)\propto \left(\eta/\eta_0\right)^{-1}$ can not be merely factored out from the integration. It seems rather difficult to take into account the consequences of this inherent relationship rigorously. Because of this we have to resort to the crucial approximation (\ref{eq16}). We have presented arguments to justify the latter. Nevertheless we have to recognize that the approximation (\ref{eq16}) introduces a hardly controllable error in our analysis. Further investigations are necessary for clarification of its validity.

Expressed mathematically the relationship between the preexponent of the effective reaction rate constant and the distribution over the activation barrier energies for enzymatic/protein reactions is formalized in the integral equation (\ref{eq30}). The latter (at the specified condition of the validity of the approximation (\ref{eq25}) or equivalently (\ref{eq16})) is valid as long as the requirement that the substrate is in excess (\ref{eq33}) is satisfied. The solution of this integral equation yields the normalized distribution over the values of the activation barrier energy (\ref{eq36}). The typical behavior of the obtained distribution over the activation barrier energies is depicted in Fig. 1 and Fig. 2. They show that depending on the combinations of the parameters $\beta$ and $q$ two cases are possible within the range of validity of our approach $q >> 1$: 1. the distribution becomes wider with the decrease of the parameter $\beta$ while the position of its maximum retains its value unchanged (see Fig. 1) and  2. the distribution becomes wider with the decrease of the parameter $\beta$ while the position of its maximum is shifted to higher values (see Fig. 2).

The equation (\ref{eq30}) resembles in its origin and structure the so-called antigen binding equation arising in immunology but differs from the latter in details. It is the Fredholm integral equation of the first kind. Such equation is known to be an example of the so-called ill-posed problem (small variations in the kernel of the integral equation may lead to large deviations in the solution). As a rule  special regularization methods are necessary for the analysis of such equations. However a solution of such equation is considered as the "dangerous" one assuming it contains sharp deflexions of the function at small changes of the variable (in other words if it is not smooth). Fig. 1 and Fig.2 testify that in our case the solution (\ref{eq36}) of the integral equation (\ref{eq30}) yields very smooth functions. Thus we conclude that the solution obtained in the present paper is "safe" as regards the ill-posed problem criterion.

The analysis of experimental data within the framework of the present approach requires information on both viscosity and temperature dependence of the reaction rate constant. That is from the experimentally observable value of the effective reaction rate constant
\[
k_2^{eff}=\frac{b}{\left(\eta/\eta_0\right)^\beta}\exp\left(-\frac {E}{k_B T}\right)
\]
we need to extract three parameters: $b$, $\beta$ and $E$. Besides the value of the reaction rate constant $k_{-1}$ for enzymatic reaction or its analog $k_{int}$ for protein reactions (see previous Sec.) is necessary for the analysis.
All these parameters  are considered as input information for the present model. The theoretically calculated value of the parameter $a$ must closely match (or at least be commensurable with) the given value $b$ extracted from experimental data. The latter requirement is the criterion for the validity of the present approach. The example of experimental data for oxygen escape from hemerythin with viscosity varied by glycerol from the paper \cite{Yed95} analyzed in the previous Sec. testifies that this criterion can in fact be satisfied for realistic situations. Regretfully these data seem to be unique as regards their completeness and sufficiency for our approach. New experimental measurements are highly desirable in this field of science especially concerning enzymatic rather than protein reactions.

The present approach retains the main idea of our previous model \cite{Sit10} and its capacity for interpretation of the experimental data on solvent viscosity dependence for enzymatic reactions. However technical implementation of the present model is different from that of \cite{Sit10}. Here we make use of the distribution over the activation barrier energies that has appreciable conceptual advantages. The latter distribution is not a mere abstract notion.
This distribution is virtually turned to a directly observable experimental function by the progress in single enzyme kinetics measurements. The distribution over the activation barrier energies obtained in the present paper is of a familiar bell-shaped form. In \cite{Sit10} we obtained the distribution over the weight with which the contribution from solvent viscosity is taken into account in the viscosity for the movement of the system along the reaction coordinate. In contrast to the present one the distribution from \cite{Sit10} is somewhat unusual. The latter has divergencies (although integrable ones, i.e., the distribution is still normalized) at both ends of the range. Thus the distribution obtained in the present paper intuitively seems to be more simple and comprehensible.
This simplification in physical picture requires increased mathematical complexity of the present model compared with that from \cite{Sit10}. The integral equation obtained in the latter paper had exact solution. For the integral equation derived in the present paper we have been able to obtain only an approximate solution. However this solution yields excellent accuracy within the physically interesting range of the parameters outlined by the requirement (\ref{eq35}). It has been verified by numerical integration.
We can formulate the range of the validity of our distribution (\ref{eq36}) as follows: the latter works better when the requirement (\ref{eq35}) $q >> 1$ is fulfilled with more accuracy. In the case of oxygen escape from hemerythin for which we have rather small value $q \approx 15.85$ the requirement (\ref{eq35}) is satisfied with some strain. However
from Fig.5 we see that even in this case we obtain satisfactory approximation for the solution of the integral equation (\ref{eq30}). A small value of $q$ for this protein reaction is the result of rather low effective activation barrier energy $H_{esc}=13\ kJ \cdot mol^{-1}$ at $278\ K$ \cite{Yed95} that yields $E/\left(k_B T\right)\approx 5.63$. For most enzymatic reactions the effective activation barrier energies are much higher so that the ratio of $E/\left(k_B T\right)$ attains typical values $15 \div 20$ at room temperature. From Fig. 3 we see that for such cases the parameter $q$ attains very large values up to $10^3 \div 10^4$. Thus we conclude that for these reactions the requirement (\ref{eq35}) is fulfilled very well. Therefore we conclude that for most enzymatic reactions our distribution (\ref{eq36}) provides approximation to the solution of the integral equation (\ref{eq30}) with excellent accuracy. Therewith it should be remembered that the mentioned above reservations on the validity of the equation (30) remain in force.

The obtained bell-shaped form of the distribution  over the activation barrier energies is in qualitative agreement with that of the distribution over the reaction rate constant obtained in single enzyme experiments \cite{Ris08}. However the search for any quantitative correspondence seems to be meaningless and premature.
The single enzyme kinetics measurements and the studies of solvent viscosity dependence for enzymatic reactions have been developed independently and no objects of mutual interest have been explored by now. That is why direct comparison of the distribution obtained from our analysis with those extracted from single enzymes kinetics is unfeasible.
It seems highly desirable to explore an enzyme for which on the one hand single enzyme kinetics measurements would be feasible and on the other hand sufficiently extensive experimental data for solvent viscosity effect on its reaction rate constant would be available. The data on such object would provide the possibility for the direct and crucial experimental verification of the predictions of the present model. In this regard the investigations of solvent viscosity dependence for cholesterol oxidase (which is the main object for single enzyme kinetics measurements \cite{Lu98}) and that for $\beta$-galactosidase (for which in \cite{Ris08} the distribution over the reaction rate constant is obtained) seem to be of primary interest. The experimental data on solvent viscosity dependence for these enzymes
of the same completeness as those for oxygen escape from hemerythin
would be invaluable for quantitative verification of the present model.

We conclude that there is the inherent relationship between the distribution over the activation barrier energies and the preexponent of the effective reaction rate constant for the solution of enzymes. Our model yields simple interpretation and the quantitative description of the available experimental data on solvent viscosity dependence for enzymatic/protein reactions. The approach is in conceptual coherence with the modern trend stimulated by single enzyme kinetics.
\\

Acknowledgements.  The author is grateful to Dr. Yu.F. Zuev for
helpful discussions. The work was supported by the grant from RFBR and
the programme "Molecular and Cellular Biology" of RAS.

\newpage

\begin{figure}
\begin{center}
\includegraphics* [width=\textwidth] {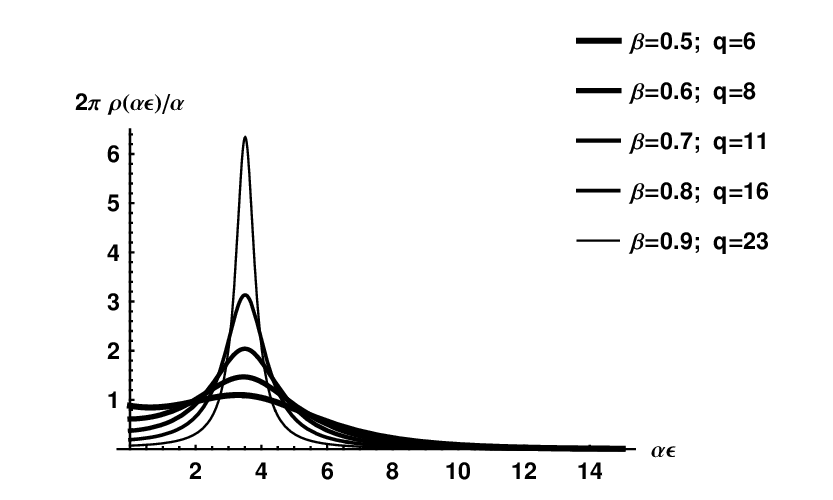}
\end{center}
\caption{Distribution $\rho (\epsilon)$ over the activation barrier energies $\epsilon$ (eq. (\ref{eq36})) at increasing values of the fractional power exponent $\beta$ and different values of the parameter $q$: $\beta =0.5$, $q=6$ (thick line); $\beta =0.6$, $q=8$; $\beta =0.7$, $q=11$ ; $\beta =0.8$, $q=16$;  $\beta =0.9$, $q=23$ (thin line).} \label{Fig.1}
\end{figure}

\clearpage
\begin{figure}
\begin{center}
\includegraphics* [width=\textwidth] {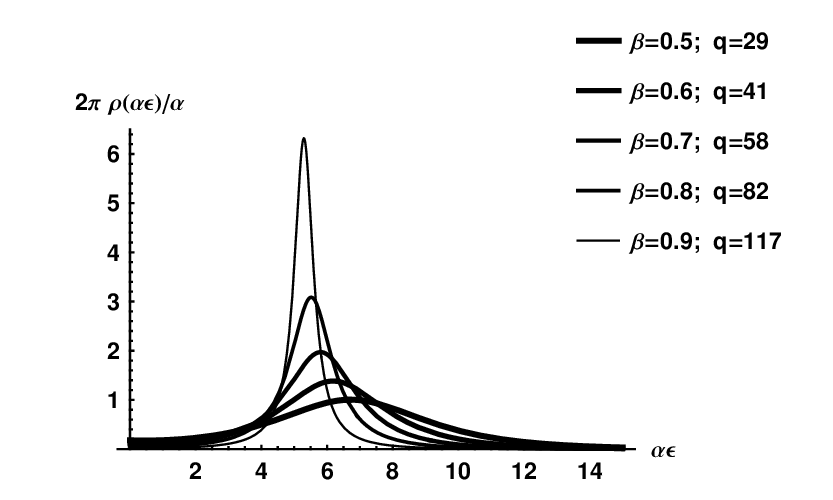}
\end{center}
\caption{Distribution $\rho (\epsilon)$ over the activation barrier energies $\epsilon$ (eq. (\ref{eq36})) at increasing values of the fractional power exponent $\beta$ and different values of the parameter $q$: $\beta =0.5$, $q=29$ (thick line); $\beta =0.6$, $q=41$; $\beta =0.7$, $q=58$ ; $\beta =0.8$, $q=82$;  $\beta =0.9$, $q=117$ (thin line).}
\label{Fig.2}
\end{figure}

\clearpage
\begin{figure}
\begin{center}
\includegraphics* [width=\textwidth] {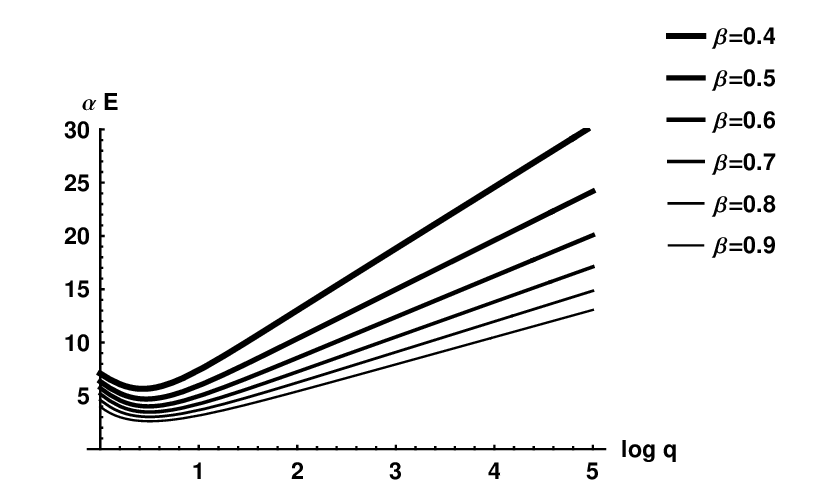}
\end{center}
\caption{Relationship between the parameters $q$ and $E$ (eq. (\ref{eq38})) at increasing values of the fractional power exponent $\beta$: $\beta =0.4$ (thick line); $\beta =0.5$; $\beta =0.6$; $\beta =0.7$; $\beta =0.8$; $\beta =0.9$ (thin line).}
\end{figure}

\clearpage
\begin{figure}
\begin{center}
\includegraphics* [width=\textwidth] {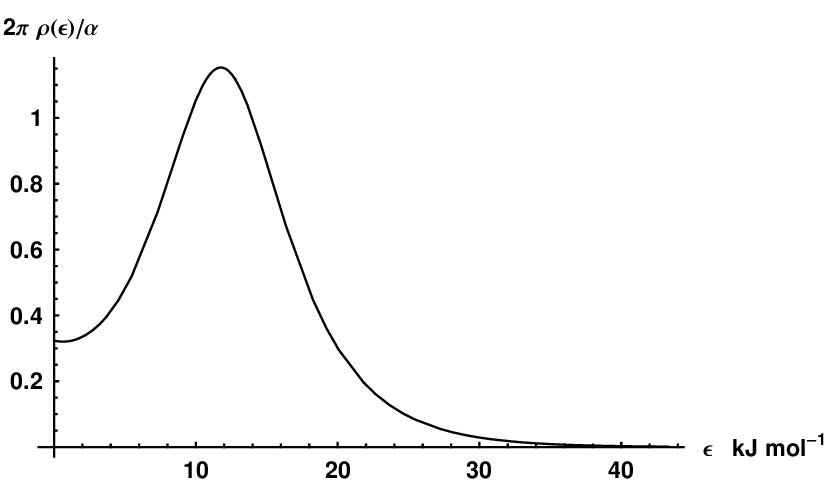}
\end{center}
\caption{Distribution $\rho (\epsilon)$ over the activation barrier energies $\epsilon$ (eq. (\ref{eq36})) for the realistic case of experimental data on oxygen escape from hemerythin from the paper (\ref{eq30}). The values of the parameters are: $\beta=0.54$; $T=278\ K$ ($\alpha=1/(k_B T)\approx 2.6\cdot 10^{13}\ erg^{-1}=2.6\cdot 10^{23}\ kJ^{-1}$); $q=15.85$.}
\label{Fig.4}
\end{figure}

\clearpage
\begin{figure}
\begin{center}
\includegraphics* [width=\textwidth] {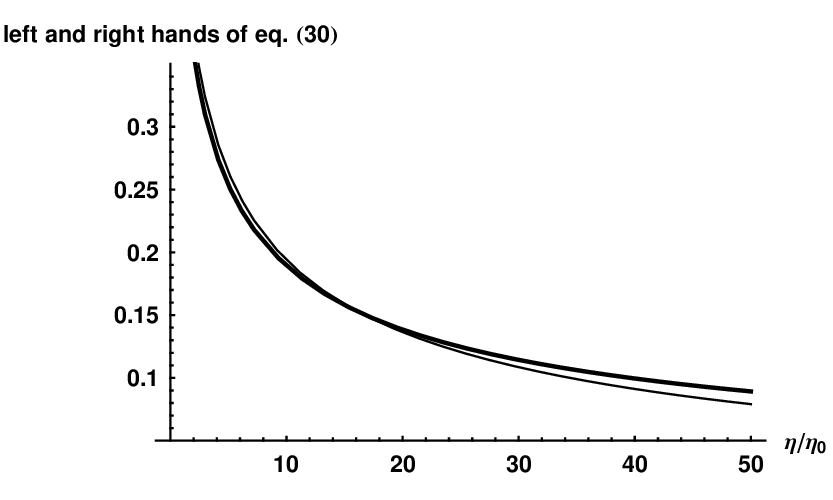}
\end{center}
\caption{Left (thin line) and right (thick line) hands of eq.(\ref{eq30}) with the distribution $\rho (\epsilon)$ over the activation barrier energies $\epsilon$ (eq.(\ref{eq36})) for the realistic case of experimental data on oxygen escape from hemerythin from the paper (\ref{eq30}). The values of the parameters are: $\beta=0.54$; $T=278\ K$ ($\alpha=1/(k_B T)\approx 2.6\cdot 10^{13}\ erg^{-1}=2.6\cdot 10^{23}\ kJ^{-1}$); $q=15.85$.}
\label{Fig.5}
\end{figure}

\newpage
\begin{thebibliography}{00}
\bibitem{Aus75}
Austin, R.H., Beeson, K.W., Eisenstein, L., Frauenfelder, H.,
and Gunsalus, I.C., 1975, Biochemistry, 14, 5355.
\bibitem{Ans85}
Ansari, A., Berendzen, J., Bowne, S.F., Frauenfelder, H., Iben, I.E.T.,
Sauke, T.B., Shyamsunder, E., and Young, R.D. 1985, Proc. Natl. Acad. Sci. USA,
82, 5000.
\bibitem{Fra91}
Frauenfelder, H., Sligar, S.G., and Wolynes, P.G. 1991,
Science, 254, 1598.
\bibitem{Fra01}
Frauenfelder, H., McMahon, B.H., Austin, R.H., Chu, K., and Groves, J.T., 2001,
Proc. Natl. Acad. Sci. USA, 98, 2370.
\bibitem{Xue95}
Xue, Q., and Yeung, E.S. 1995, Nature, 373, 681.
\bibitem{Xue06}
Xue, X., Liu, F., and Ou-Yang, Z. 2006, Phys. Rev E, 74, 030902(R).
\bibitem{Lu98}
Lu, H.P., Xun, L., and Xie, X.S. 1998, Science, 282, 1877.
\bibitem{Kou05}
Kou, S.C., Cherayil, B.J., Min, W., English, B.P., and Xie, X.S. 2005,
J. Phys. Chem. B, 109, 19068.
\bibitem{Min05}
Min, W., English, B.P., Luo, G., Cherayil, B.J., Kou, S.C., and Xie, X.S. 2005,
Acc. Chem. Res., 38, 923.
\bibitem{Ris08}
Rissin, D.M., Gorris, H.H., and Walt, D.R. 2008,
J. Am. Chem. Soc., 130, 5349.
\bibitem{Pra07}
Prakash, M.K., and Marcus, R.A. 2007, Proc. Natl. Acad. Sci. USA, 104, 15982.
\bibitem{War78}
Warshel, A. 1978, Proc. Natl. Acad. Sci. USA, 75, 5250.
\bibitem{War84}
Warshel, A. 1984, Proc. Natl. Acad. Sci. USA, 81, 444.
\bibitem{Ols06}
Olsson, M.H.M., Parson, W.W., and Warshel, A. 2006,
 Chem. Rev., 106, 1737.
\bibitem{Ols006}
Olsson, M.H.M., Mavri, J., and Warshel, A. 2006,
Phil. Trans. R. Soc. B, 361, 1417.
\bibitem{War06}
Warshel, A., Sharma, P.K., Kato, M., Xiang, Y., Liu, H., and Olsson, M.H.M. 2006,
Chem. Rev., 106, 3210.
\bibitem{Fra99}
Frauenfelder, H., Wolynes, P.G., and Austin, R.H. 1999,
Rev. Mod. Phys., 71, 419.
\bibitem{Gav78}
Gavish, B. 1978, Biophys. Struc. Mech., 4, 37.
\bibitem{Gav79}
Gavish, B., and Werber, M.M. 1979, Biochemistry, 18, 1269.
\bibitem{Bee80}
Beece, D., Eisenstein, L., Frauenfelder, H., Good, D.,
Marden, M.C., Reinisch, L., Reynolds, A.H., Sorensen, L.B.,
and Yue, K.T. 1980, Biochemistry, 19, 5147.
\bibitem{Gav80}
Gavish, B. 1980, Phys.Rev.Lett., 44, 1160.
\bibitem{Dos83}
Doster, W. 1983, Biophys. Chem., 17, 97.
\bibitem{Gav86}
Gavish, B. in: 1986,  The fluctuating enzyme, C.R. Welsh (Ed.), Wiley, N.Y., p. 263.
\bibitem{Fra88}
Frauenfelder, H., Parak, F., and Young, R.D. 1988, Ann.Rev.Biophys.Chem., 17, 451.
\bibitem{Dem89}
Demchenco, A.P., Rusyn, C.I., and Saburova, E.A. 1989,
Biochem. et Biophys. Acta, 998, 196.
\bibitem{Ng91}
Ng, K., and Rosenberg, A. 1991, Biophys Chem., 39, 57.
\bibitem{Ng911}
Ng, K., and Rosenberg, A. 1992, Biophys Chem., 41, 289.
\bibitem{Gav94}
Gavish, B., and Yedgar, S. in: 1995, Protein-solvent interactions,
R.B. Gregory (Ed.), Dekker, N.Y., p. 343.
\bibitem{Dos94}
Doster, W., Kleinert, Th., Post, F., and Settles, M. in: 1995, Protein-solvent interactions,
R.B. Gregory (Ed.), Dekker, N.Y., p. 375.
\bibitem{Yed95}
Yedgar, S., Tetreau, C., Gavish, B., and Lavalette, D. 1995, Biophys. J., 68, 665.
\bibitem{Bar95}
Barshtein, G., Almagor, A., Yedgar, S., and Gavish, B. 1995, Phys. Rev. E, 52, 555.
\bibitem{Oh97}
Oh-oka, H., Iwaki, M., and Itoh, S. 1997, Biochemistry, 36, 9267.
\bibitem{Kle98}
Kleinert, Th., Doster, W., Leyser, H., Petry, W.,
Schwarz, V., and Settles, M. 1998, Biochemistry, 37, 717.
\bibitem{Lav99}
Lavalette, D., T\'etreau, C., Tourbez, M., and Blouquit, Y. 1999,
Biophys. J., 76, 2744.
\bibitem{Uri03}
Uribe, S., and Sampedro, J.G. 2003, Biol. Proced. Online, 5, 108.
\bibitem{Lav06}
Lavalette, D., Hink, M.A., Tourbez, M.,
T\'etreau, C., and Visser, A.J. 2006, Eur. Biophys. J., 35, 517.
\bibitem{Pab04}
Pabit, S.A., Roder, H., and Hagen, S.J. 2004, Biochemistry, 43, 12532.
\bibitem{Fra06}
Frauenfelder, H., Fenimore, P.W., Chen, G., and McMahon, B.H. 2006,
Proc. Natl. Acad. Sci. USA, 103, 15469.
\bibitem{Kum08}
Kumar, R., and Bhuyan, A.K. 2008, J. Phys. Chem. B, 112, 12549.
\bibitem{Jas01}
Jas, G.S., Eaton, W.A., and Hofrichter, J. 2001, J. Phys. Chem. B, 105, 261.
\bibitem{Gao06}
Gao, J., Ma, S., Major, D.T., Nam, K., Pu, J., and Truhlar, D.G. 2006,
Chem. Rev., 106, 3188.
\bibitem{Han90}
H\"anggi, P., Talkner, P., and Borkovec, M. 1990, Rev.Mod.Phys., 62, 251.
\bibitem{Gro80}
Grote, R.F., and Hynes, J.T. 1980, J.Chem.Phys., 73, 2715.
\bibitem{Zwa92}
Zwanzig, R. 1992, J. Chem. Phys., 97, 3587.
\bibitem{Dry10}
Dryga, A., and Warshel, A. 2010, J. Phys. Chem. B, 114, 12720.
\bibitem{Sch88}
Schlitter, J. 1988, Chem. Phys., 120, 187.
\bibitem{Sit08}
Sitnitsky, A.E. 2008, Physica A, 387, 5483.
\bibitem{Sit10}
Sitnitsky, A.E. 2010, Chem. Phys., 369, 37.
\bibitem{Dix79}
Dixon, M., and Webb, E.C. 1979, Enzymes, 3-d edition, Longman, Ch.4, p.93.
\end{thebibliography}
\end{document}